\documentclass{cpcauth}

\usepackage{epsfig}

\usepackage{amssymb}

\begin{document}

\begin{frontmatter}



\title{Low-Temperature Long-Time Simulations of Ising Ferromagnets 
using the Monte Carlo with Absorbing Markov Chains method}  


\author{M.~A.\ Novotny}
\address{Department of Physics and Astronomy, Mississippi 
State University, Mississippi State, Mississippi 39762, USA}

\begin{abstract}
The Monte Carlo with Absorbing Markov Chains (MCAMC) method is 
introduced.  This method is a generalization of the rejection-free 
method known as the $n$-fold way.  The MCAMC algorithm is applied 
to the study of the very low-temperature properties of the 
lifetime of the metastable state of Ising ferromagnets.  
This is done both for 
square-lattice and cubic-lattice nearest-neighbor models.  
Comparison is made with exact low-temperature predictions, in 
particular the low-temperature predictions that the metastable 
lifetime is discontinuous at particular values of the field.  
This discontinuity for the square lattice is not 
seen in finite-temperatures studies.  
For the cubic lattice, it is shown 
that these \lq exact predictions' are incorrect near the fields 
where there are discontinuities.  The low-temperature formula 
must be modified and the corrected low-temperature predictions are 
not discontinuous in the energy of the nucleating droplet.   
\end{abstract}

\begin{keyword}
Advanced algorithms \sep Dynamic Monte Carlo \sep lifetimes \sep
metastable \sep prefactor

\PACS 05.10.-a \sep 05.10.Ln
\end{keyword}
\end{frontmatter}

\section{Introduction}

One of the most difficult problems facing simulations in science and 
mathematics is to be able to simulate time and length scales comparable to 
those found in nature, those required for engineering applications, and 
those accessible to experiments.  In this paper a method to obtain 
extremely long simulations for discrete systems is reviewed.  
This method, which merges the ideas of absorbing Markov Chains with 
those of kinetic Monte Carlo simulations, is called the 
MCAMC algorithm (Monte Carlo with Absorbing Markov Chains) 
\cite{MAN95,MANUGA2,MANreview}.  It expands on the 
ideas put forward by Bortz, Kalos, and Lebowitz \cite{BKT} in a 
method called the $n$-fold way.  The MCAMC algorithm 
with a single transient state (which corresponds to the 
current state of the configuration) is the discrete-time 
version of the $n$-fold way algorithm \cite{MANCinP}.  

As an application requiring long-time simulations, consider the 
problem of thermal reversal of nanoscale ferromagnetic grains.  
For highly anisotropic materials each region of the ferromagnet 
may be considered to have two states corresponding to two discrete 
directions.  This gives an Ising model as an 
approximate Hamiltonian for ferromagnetic nanoparticles \cite{Howard1}.  
Starting from a quantum Hamiltonian of spin $1\over 2$ particles 
interacting with a heat bath, it is 
possible to derive in certain limits a physical dynamic for the 
model \cite{Martin,KParkCCP}.  This dynamic has two parts: 
first a spin is chosen uniformly from all the spins, then the 
spin is flipped with some probability $p_{\rm flip}$ that depends 
on the temperature $T$ and the applied magnetic field $H$.  
One algorithmic step is called a Monte Carlo step (mcs) and corresponds to 
one spin-flip attempt.  For a system with $N$ Ising spins, the 
simulation time is often given in Monte Carlo steps per spin (MCSS), 
which corresponds to $N$ mcs.  

With coupling to a fermionic heat bath \cite{Martin}, 
the probability $p_{\rm flip}$ is the Glauber transition probability 
\cite{Glauber}.  
In this case 
$
p_{\rm flip} = {\rm e}^{-\beta E_{\rm new}} 
/ [{\rm e}^{-\beta E_{\rm new}}+{\rm e}^{-\beta E_{\rm old}}]
$
where $\beta=T^{-1}$ (with Bolztmann's constant set to one), 
$E_{\rm new}$ is the energy of the configuration with the 
Ising spin flipped and $E_{\rm old}$ is the energy of the 
current spin configuration of the system.  
A different probability (which also satisfies detailed balance) 
for coupling to a $d$-dimensional bath of phonons 
has been recently derived \cite{KParkCCP}.  

The reason long-time simulations are required is now readily apparent.  
The underlying algorithmic step corresponds to an inverse phonon 
frequency, about $10^{-13}$~seconds.  The time scale for simulation 
must be on the order of years to decades for simulations 
applicable to magnetic recording --- basically simulations at least as 
long as data integrity for a written bit of information.  For 
paleomagnetism simulations, time periods of millions of years 
must be achievable.  
The typical clock speed of a computer is $10^{-9}$~seconds.  
Consequently, faster-than-real-time simulations 
are required for realistic feasible simulations.  
Note that these faster-than-real-time algorithms {\it cannot\/} 
change the dynamic which has been derived starting from the 
quantum Hamiltonian.  Hence advanced algorithms that are used 
in static simulations, such as reweighting techniques, 
the Swendsen-Wang algorithm, multicanonical methods, or simulated 
tempering \cite{MANreview} cannot be used --- they would change the 
underlying dynamic.  Rather, the physical dynamic described above must 
be implemented on the computer in a more intelligent fashion than a 
brute-force method.  

This paper describes only the MCAMC algorithm and results obtained 
from it for low-temperature 
Ising simulations.  Two other recent advances using 
rejection-free methods 
should be mentioned \cite{MANreview}.  
One is to use massively parallel computers in such simulations 
\cite{MANreview,GKnfold,GyorgyPRL,GyorgyUGA}.  
Another is to generalize the rejection-free algorithms 
to continuous systems, such as Heisenberg spin systems 
\cite{MunozUGA}.

Sec.~2 describes the MCAMC algorithm for the 
simple cubic (sc) nearest-neighbor (nn) Ising model.  
The generalization to other discrete systems is straightforward.  
In Sec.~3 low-temperature predictions for the Ising model are 
reviewed.  Sections 4 and 5 present results for low-temperature 
metastable lifetimes in $d$$=$$2$ and 
$d$$=$$3$.  Sec.~5 contains conclusions.  

\section{The MCAMC Algorithm} 

We simulate the nn Ising model with Hamiltonian 
$
{\mathcal H} = - J \sum_{\langle i,j\rangle} \sigma_i \sigma_j 
- H \sum_i \sigma_i 
$
with $J$$>$$0$ the ferromagnetic coupling constant, 
the first sum running over all nn pairs, and 
the second sum running over all $N$ spins.  
We consider only the case of periodic boundary conditions.  
The simulation starts with all spins $\sigma_i=1$ (all spins up), and 
a field $H<0$ (external field down) is applied.  The magnetization 
is $M=\sum_i^N \sigma_i$.  We measure the time $\tau$ 
to go from the starting state to a state with $M=0$.  
This simulation is repeated many times (for this paper 1000 times) 
for the same $T$ and $H$ values to obtain 
the average lifetime $\langle\tau\rangle$ of the metastable state.  

For the sc lattice for every configuration each spin can be 
classified to be in one of $14$ possible classes.  
(For the square-lattice Ising model there are $10$ classes.)  
The spin is either up or down, and we label the first 7 classes 
with spin up and classes 8 through 14 with spin down.  
The other factor determining the class of a spin is the number of 
nn spins that are up, which can be any integer from 
zero to 6.  Let class~1 have 6 nn spins up, class~2 5 nn spins up, 
etc.  Class 8 has 6 nn spins up, $\cdots$, class 14 has 0 
nn spins up.  Let $n_i$ be the number of spin in class $i$.  
Then $N = \sum_{i=1}^{14}n_i$.  
Let $p_i$ be the probability of flipping a spin in class $i$, 
given that the spin was chosen during the first part of the dynamic 
algorithm.  
Then the probability of flipping any of the spins in class $i$ in 
one mcs is $n_i p_i / N$ since $n_i/N$ is the probability of 
choosing a spin from class $i$.  

In order to exit from the 
current spin configuration, one of the spins in one of the $14$ 
classes must be flipped.  
Define $Q_i={1\over N} \sum_{j=1}^i n_j p_j$ 
for each of the $14$ spin classes, and  $Q_0=0$.  
Let $\lfloor\cdot\rfloor$ denote the integer part of a number.  
The discrete $n$-fold way algorithm has three steps.  
First, the time $m$ to exit from the current configuration (in 
unit of mcs) is determined by 
\begin{equation}
m = \left\lfloor
{{\ln({\bar r})}\over{\ln(1-Q_{14})}}
\right\rfloor + 1
\end{equation}
with ${\bar r}$ a random number uniformly distributed between 
zero and one.  
Then using a different uniformly distributed random 
number ${\hat r}$ the spin class $k$ is chosen such that it 
satisfies 
$
Q_{k-1}\le {\hat r} Q_{14} < Q_k  .
$
Finally, using a third uniformly distributed random number, one of 
the $n_k$ spins in class $k$ is chosen, and this spin is 
flipped.  Of course, the flipped spin changes its class, as do 
its 6 nn spins.  This means that the 
numbers $n_i$ change after one algorithmic step.  

The first two steps of the discrete $n$-fold way algorithm 
correspond to having an absorbing Markov chain with 
$14$ states in the recurrent matrix with elements given by 
the probability to exit from the current spin configuration by 
choosing and flipping a spin in class $i$, namely 
$n_i p_i / N$.  The transient subspace of the absorbing Markov 
matrix is a scalar ($s=1$) given by 
the matrix ${\bf T}_{s=1}=N(1-Q_{14})$.  
The time in the $s=1$ MCAMC simulation corresponds to the time 
required to leave the current spin configuration.  At low temperatures 
this time may be extremely long (if $Q_{14}$ is very small), and the 
algorithmic speed-up obtained in the simulation can be many 
orders of magnitude.  

The simulation starts with all spins up.  
In one algorithmic step the $s=1$ MCAMC described above exits from 
this state with a probability $p_1$ (since all spins are up, 
$n_1=N$ for this state).  
For low temperatures and $H<2J$ the most probable thing that 
happens in the next algorithmic step is that the flipped spin is 
again chosen and the next configuration is again all spins up.  
This occurs with probability ratio 
$
{(N-7)p_1+6p_2}\over p_8
$.  
By increasing the size of the transient subspace of the 
absorbing Markov chain, it is possible to exit the absorbing 
Markov chain in a configuration with more than one spin flipped.  
For example, the absorbing Markov matrix to exit from the state 
with all spins up into a state with two spins flipped has 
two states in the transient subspace ($s=2$) with the 
transient matrix 
\begin{equation}
{\bf T}_{s=2} = {1\over N} 
\pmatrix{
x_2 & p_8 \cr
N p_1 & N(1-p_1) \cr 
}
\end{equation}
and the recurrent matrix 
\begin{equation}
{\bf R}_{s=2} = {1\over N} 
\pmatrix{
6 p_2 & (N-7)p_1 \cr
0 & 0 \cr 
}
\end{equation}
with $x_2 = N-p_8-6p_2-(N-7)p_1$.  
The ratio to exit to a state with two nn spins 
overturned to that to exit to a state with 
two non-nn spins overturned is $6 p_2/(N-7)p_1$.  
The vector representing the initial state of 
all spins up is
$
{\overrightarrow{v}}_{\rm I}^{\rm T} =
\pmatrix{
0 & 1 \cr
}
$, 
and the probability to be in each of the 
two transient states after one time step is 
${\overrightarrow{v}}_{\rm I}^{\rm T} {\bf T}_{s=2}$.  
The first transient state (lower right-hand corner of the matrix) is 
the state with all spins up, and the other state (upper left-hand 
corner of the matrix) corresponds to the $N$ states that have 
one overturned spin.  
By adding more states to the transient subspace longer 
times to exit the transient subspace are possible.  

The time $m$ (in mcs) 
to exit from the transient subspace is found 
from the solution of the equation 
\begin{equation}
{\overrightarrow{v}}_{\rm I}^{\rm T} {\bf T}^{m} {\overrightarrow{e}}
< {\bar r} \le
{\overrightarrow{v}}_{\rm I}^{\rm T} {\bf T}^{m-1} {\overrightarrow{e}}
\end{equation}
where 
${\overrightarrow{e}}$ is the vector of length $s$ with all elements 
equal to unity.  
The probability that the system ends up in a particular absorbing state is 
given by the elements of the vector
$
{\overrightarrow{v}}_{\rm I}^{\rm T} {\bf N} {\bf R}
$
with the fundamental matrix 
${\bf N}=({\bf I}-{\bf T})^{-1}$.  

To obtain very long lifetimes, a multiple precision 
package was used \cite{Bailey}.  
To keep bookkeeping overhead to a minimum, for the Ising 
simulations below, only $s=1$ MCAMC was used except when 
the spin configuration had all spins up, then higher 
$s$ MCAMC (up to $s=5$) was used.  

\section{Low-temperature Metastable Ising Predictions} 

At very low temperatures the kinetic Ising 
simulations are influenced by the discreteness of the lattice.  
This allows for the exact calculation of the saddle point as 
well as the most probable route to the saddle point \cite{NevSch}.  
For the square lattice \cite{NevSch} with 
$\ell_2=\lfloor 2J/|H|\rfloor +1$ the 
average lifetime (in units of mcs) at low temperature is given by 
$\langle\tau\rangle=A_2 \exp(\Gamma_2/T)$ with 
$\Gamma_2=8J\ell_2 -2|H|(\ell_2^2-\ell_2+1)$.  
The prefactor, $A_2$, was determined to be 
$5/4$ for $\ell_2=1$ and $3/8$ for $\ell_2=2$ \cite{MANUGA2}.  
Recently, the prefactor has been found to be 
$A_2=3/[8(\ell_2-1)]$ for all $|H|<2J$ \cite{RecPref}.  
Hence the nucleating droplet for the square lattice is a 
rectangle of overturned spins of size $\ell_2\times(\ell_2-1)$ 
with one overturned spin on a long interface.  
At low enough temperature, this result for $\langle\tau\rangle$ 
should hold everywhere, except when $2J/|H|$ is an integer.  
Note that the prefactor is discontinuous for the 
special values where $2J/|H|$ is an integer.  

For the sc lattice, the average lifetime is given by 
$\langle\tau\rangle = A_3 \exp(\Gamma_3/T)$ with 
$\ell_3=\lfloor 4J/|H|\rfloor+1$ and 
\begin{equation}
\label{Gamma3}
\Gamma_3 = 12J\ell_3^2-8J\ell_3-2|H|\ell_3^2(\ell_3-1) + \Gamma_2
\label{GAMMA3} 
\end{equation}
\cite{RecPref,CFQ1,CFQ2}.  
This result should hold when $4J/|H|$ is not an integer and 
when $\ell_3\ge2$.  The prefactor has recently been found to be 
$A_3=\left[16(\ell_3-\ell_2+1)(\ell_2-1)\right]^{-1}$
\cite{RecPref} for $\ell_3\ge2$.  
Hence the nucleating droplet is a cube with length $\ell_3$ with 
one layer removed and the square-lattice nucleating droplet placed 
on this layer.  
Note that at the special values where $4J/|H|$ is an integer 
$\Gamma_3$ is discontinuous, and in fact decreases by $4J$.  

\section{Square-lattice Results} 

\begin{center}
\begin{figure}[t]
\includegraphics[width=7cm,height=5cm]{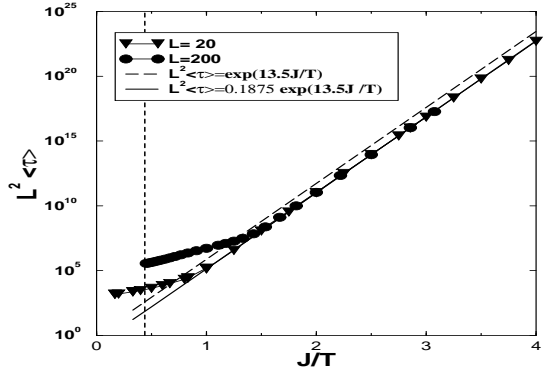}
\caption{
Mean lifetime, 
$\langle \tau \rangle$, in units of
mcs as a function of $T^{-1}$ at $|H|$$=$$3J/4$
for the square lattice using the Glauber dynamic. The symbols are 
$s$$=$3 MCAMC data. 
The dashed line is the low-temperature prediction 
with the prefactor set to one, while the heavy solid line is with 
the correct prefactor of $3/16$ \protect\cite{RecPref}.
The vertical dashed line marks $T_{\rm c}$.
}
\label{eps2d1}
\end{figure}
\end{center}

The MCAMC values obtained for the square-lattice lifetime at 
$|H|=3/4$ (so $\ell_2=3$) are shown in Fig.~\ref{eps2d1}.  Note that the 
average lifetime is very large.  For $L=20$ if one Monte Carlo 
step per spin (MCSS) corresponded to about $10^{-13}$~sec then 
the largest time corresponds to about one year.  The low temperature 
prediction with prefactor one (dashed line) does not describe the 
simulation data very well.  However, including the low temperature 
prefactor \cite{RecPref} of $3/16$ for this field value 
(heavy solid line) fits the data very nicely.  

\begin{center}
\begin{figure}[t]
\includegraphics[width=7cm,height=5cm]{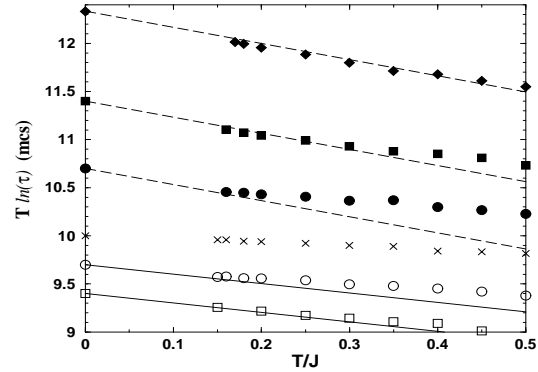}
\caption{
Average lifetime, $T \ln(\langle \tau \rangle)$, in units of 
$J$ and $\langle\tau\rangle$ in units of mcs, 
as a function of $T$ for a $L=24$ 
square lattice using the Glauber dynamic. The 
symbols at finite $T$ are 
$s$$=$3 MCAMC data. 
The values of $|H|/J$ used, reading from the top, are 
$5/6$, 0.90, 0.95, 1.00, 1.05, and 1.10.  
The symbols at $T=0$ are the exact values 
of $\Gamma_2$, and 
the lines use the exact prefactor $A_2$.  The prefactor is 
$3/8$ for $\ell_2=2$ ($J<|H|<2J$) and 
$3/16$ for $\ell_2=3$ ($2J/3<|H|<J$).  The prefactor at $|H|=1$ is 
not known.
}
\label{eps2d2}
\end{figure}
\end{center}

Figure~\ref{eps2d2} presents results for 
$T \ln\left (\langle\tau\rangle\right )$ which, with $\langle\tau\rangle$ 
in units of mcs, should equal 
$\Gamma_2 + T \ln(A_2)$.  These are 
for $L=24$ for various $|H|$ values.  
The intercept is the exact low-temperature prediction, 
$\Gamma_2$, 
and 
the linear slope is the exact low-temperature prefactor, $A_2$.  
These data indicate that both the value of $\Gamma_2$ and the 
prefactor agree at low enough temperatures with the simulation data.  
However, exactly how low is low enough to agree with the data depends 
on the value of $|H|$.  In particular, for values far from the 
special value where $\ell_2$ changes, the MCAMC values agree with the 
low-temperature predictions even for temperatures as large as 
$T=J/2$ (note $T_c\approx 2.26 J$).  However, closer to the 
special value where $\ell_2$ changes much lower temperatures are needed 
before agreement with the low-temperature predictions are seen.  
Hence the discontinuity in $\langle\tau\rangle$ due to the 
discontinuous prefactor is {\it not\/} seen in the simulation data.  

\section{Cubic-lattice Results} 

\begin{center}
\begin{figure}[t]
\includegraphics[width=7cm,height=5cm]{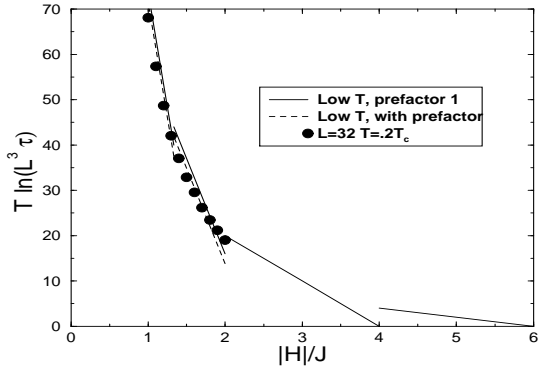}
\caption{
The average lifetime, 
$\langle \tau \rangle$, in units of
Monte Carlo steps (mcs) as a function of $|H|$ at $T=0.2 T_{\rm c}$
for the simple cubic lattice using the Glauber dynamic. The symbols are 
both MCAMC data using up to 3 transient states and 
projective dynamics simulations \protect\cite{MiroPRL} . 
The heavy solid line is the low-temperature prediction 
with the prefactor set to one, while the dashed line segments are with 
the correct prefactor \protect\cite{RecPref}.
}
\label{eps3d1}
\end{figure}
\end{center}

Figure~\ref{eps3d1} shows results for the sc lattice.  
The MCAMC and projective dynamics \cite{MiroPRL} results 
both agree with each 
other, and away from the values of $|H|$ where $\ell_3$ change 
they agree with the low-temperature prediction with the predicted 
prefactor \cite{RecPref}.  At these 
special $|H|$ values $\Gamma_3$ decreases by 
$4J$.  This would predict that near these values as $|H|$ decreases 
the lifetime of the metastable state would decrease.  This is not 
physically intuitive, and the simulation data do show evidence of 
the predicted discontinuity for $T=0.2 T_{\rm c}$.  

\begin{center}
\begin{figure}[t]
\includegraphics[width=7cm,height=5cm]{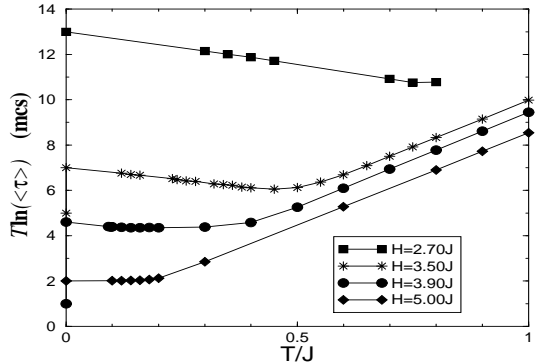}
\caption{
For the simple cubic lattice with Glauber dynamics, 
$T \ln\left(\langle \tau \rangle\right)$ 
is shown for several values of $|H|$.  
The symbols at finite $T$ are 
MCAMC data using up to $s=4$.  
For $|H|=2.7J$ below about $T=0.7J$ the MCAMC results seem to agree 
with a linear approach to the exact prediction, indicated by 
the same symbol at $T=0$.  The same is also true for the results at 
$|H|=5J$ where the $T=0$ intercept is equal to~2.  However, for 
both $|H|=3.5J$ and $|H|=3.9J$ the low-temperature 
predictions of Eq.~\protect\ref{GAMMA3} are 
{\it not\/} valid and do {\it not\/} fit the MCAMC results.  
These invalid results are shown by the lower corresponding 
symbols at $T=0$.  Rather, the MCAMC results tend toward the 
corrected predictions given by the higher corresponding $T=0$ symbols.  
}
\label{eps3d2}
\end{figure}
\end{center}

Figure~\ref{eps3d2} shows results of MCAMC simulations of the 
sc Ising model at low temperatures for various strong fields.  
No prefactors for these values (since $|H|>2J$) are predicted by 
\cite{RecPref}.  For $4J<|H|<6J$ the prefactor can be 
evaluated using absorbing Markov chains, and is $7/6$.  
This result is consistent with the slope for $T<0.2J$.  

The most striking result shown in Fig.~\ref{eps3d2} is that for 
both $|H|=3.5J$ and $|H|=3.9J$ the MCAMC results do {\it not\/} 
agree with the low-temperature predictions 
\cite{RecPref,CFQ1,CFQ2}.  These are shown by the 
lowest corresponding symbols at $T=0$.  
An analysis of the path taken by the nucleating droplet shows that 
there is a higher saddle point than that predicted by the 
low-temperature results of Eq.~\ref{GAMMA3}.  
The value expected at this higher saddle 
point is shown by the higher corresponding symbols at $T=0$.  
For $3J\le|H|\le4J$ this saddle has energy 
$28J-6|H|$, corresponding to an L-shaped droplet with three overturned 
spins.  
The region of $|H|$ for which there is a higher saddle point 
shrinks with increasing $\ell_3$, but is always finite near any 
non-zero value of $|H|$.  This higher saddle point 
removes the discontinuity in $\Gamma_3$ near the values of $|H|$ 
where $\ell_3$ changes.  In fact, $\Gamma_3$ is continuous for 
the corrected saddle, just as $\Gamma_2$ is continuous.  
The discontinuity in $\langle\tau\rangle$ due to a 
discontinous prefactor has not yet been supported by the MCAMC 
data for the sc lattice.  One might expect the same 
behavior as seen in the square lattice, which keeps the discontinuity 
from being seen at any finite temperature.  

\section{Conclusions} 

The Monte Carlo with Absorbing Markov Chains (MCAMC) algorithm 
for long-time simulations of dynamics was described.  
The description was for the sc Ising ferromagnet, 
but can be generalized to other discrete systems.  The 
$s=1$ MCAMC algorithm corresponds to the $n$-fold way 
algorithm \cite{BKT} in discrete time \cite{MANCinP}.  
In many cases the $s=1$ MCAMC can require many {\it orders of magnitude\/} 
less computer time than conventional dynamic simulations to 
perform the same calculation.  The 
$s>1$ MCAMC can require many {\it orders of magnitude\/} less 
computer time than the $s=1$ MCAMC.  These exponential decreases  
in computer time are accomplished {\it without\/} changing the 
underlying dynamic.  

The MCAMC algorithm has been applied to check the low-temperature 
predictions for metastable decay in both the square and sc 
Ising ferromagnet.  In particular, the predicted discontinuity in 
the average lifetime at the values of $|H|$ where the size of the 
droplet at the saddle point changes discontinuously was investigated.  
It was found that for the square lattice, where the discontinuity is 
only in the prefactor, a finite temperature simulation sees no 
evidence of this prefactor discontinuity.  Instead, the 
temperature at which the simulations must be performed to see the 
low-temperature predictions decreases as the discontinuity is 
approached.  For the sc lattice a discontinuity in 
both the prefactor and in the energy of the nucleating droplet 
was predicted \cite{RecPref,CFQ1,CFQ2}.  
No evidence for these discontinuities were found 
in the simulation data.  The `exact formula' for the cubic lattice 
nucleating droplet, Eq.~\ref{GAMMA3}, 
was rather found to be incorrect near the 
discontinuities.  The corrected exact formula has no discontinuity in 
the energy of the nucleating droplet.  
\\
\noindent{\textbf{Acknowledgements}} \\
Special thanks to M.\ Kolesik for allowing inclusion of 
unpublished projective dynamic data.  
Partially funded by NSF DMR-9871455.  
Supercomputer time provided by the DOE through NERSC.

\end{document}